# $\theta$-Tunable Photoluminescence from Interlayer Excitons in Twisted Bilayer Graphene

Hiral Patel[1], Lujie Huang[2], Cheol-Joo Kim[3], Jiwoong Park[2] and Matt W. Graham[1]

[1.] *Department of Physics, Oregon State University, Corvalis, OR USA*

[2.] *Department of Chemistry, Institute of Molecular Engineering, University of Chicago, Chicago, IL USA*

[3.] *Department of Chemical Engineering, Pohang University of Science and Technology, Pohang, Korea*

Using resonant 2-photon excitation of interlayer electrons in twisted bilayer graphene ($t$BLG), we resolve photoluminescence (PL) that tunes spectrally with stacking angle, $\theta$. This weak signal is 4-5× larger than the non-resonant background, and is emitted from the interlayer band anti-crossing regions traditionally associated with van Hove singularity resonances. However, our observation of resonant PL emission with delayed $\sim 1$ ps electronic thermalization suggests interlayer carriers may instead form bound-excitons. Using both the 2-photon PL and intraband transient absorption spectra, we observe bright and dark state peak-splitting associated with an interlayer exciton binding energy ranging from 0.5 to 0.7 eV for $\theta = 8^o$ to $17^o$. These results support theoretical models showing interlayer excitons in $t$BLG are stabilized by a vanishing exciton-coupling strength to the metallic continuum states. This unexpected dual metal-exciton optical property of $t$BLG suggests possible $\theta$-tuneable control over carrier thermalization, extraction and emission in optical graphene-based devices.

The constrained overlap of interlayer orbitals in twisted bilayer graphene ($t$BLG) produce optical resonances that tune monotonically with layer stacking angle, $\theta$, enhancing bilayer graphene absorption by up to ∼25% on resonance.[1–6] Currently, there are two prevailing mechanisms that explain resonant absorption in $t$BLG; the hot electron van Hove singularity (vHs) model, and the formation of bound interlayer excitons.[1, 7] In Fig. 1b$i$ we sketch the vHs model with two degenerate optical transitions labeled $X_{13}$, and $X_{24}$ occurring between band anticrossing regions of the interlayer bandstructure overlap. Recently, these flattened anticrossing regions of the $t$BLG band structure have shown transformative many-body electronic physics such as exciton effects, $\theta$-dependent superconductivity and metal-insulator transitions.[7–9] Here, we report $\theta$-tunable PL emitted from theses anticrossing regions under ambient conditions, and measure 1- and 2-photon peak-splitting consistent with stable exciton formation.

Near these band anticrossing regions, theory predicts strongly enhanced electron-hole interactions from a special interlayer orbital mixing process of degenerate states.[7] Fig. 1b$ii$ depicts this bound-exciton model after band renormalization about the vHs to give Fano resonance transitions, $X_{13}$ and $X_{24}$. Symmetric and antisymmetric mixing of these two highly-degenerate Fano resonances yields excitonic interlayer states; namely an optically bright state, $X_S$ and a dark strongly-bound state, $X_A$. Simulations predict that the lower-lying exciton state, $X_A$ is stabilized by a vanishing coupling term ($H_k$, see Fig. 1b$i$) between the exciton, and lower-lying metallic graphene states.[7] We posit the dark $X_A$ state and higher lying optically dark transitions are observable through 2-photon selection rules similar to the hydrogenic exciton models used for analogous materials like the semiconducting carbon nanotubes.[10–12]

To delineate the vHs model from the bound-exciton model, we analyze 2-photon photoluminescence (PL) and transient absorption (TA) spectra. In the vHs model, resonant enhancement of $\theta$-tunable PL is not expected because interlayer hot electrons are strongly coupled to continuum states, and will thermalize rapidly (∼10 fs) by impulsive electron-electron scattering (see Fig. 1b).[13] Instead, our observation of resonant PL emission from the bright, $X_S$ state of $t$BLG is more consistent with GW-BSE simulations by Liang *et al.* that show interlayer electron-hole pairs are both strongly bound and stable excitons.[7] Lower bounds for the $t$BLG exciton binding energy, $E_b$ can be estimated using both 2-photon PL excitation (PLE) spectra and intraband ESA transitions to higher lying states near the $e$-$h$ continuum. The resulting spectral peak-splitting provide the first such $E_b$ estimates of interlayer electronic states for many single domains of $t$BLG.

**Results:** While hot-electron PL from graphene and $t$BLG have been reported, this is the first report of resonant PL in graphene emitted from the anticrossing regions.[14, 15] To observe this weak emission, the state must be resonantly prepared, and 2-photon excitation was necessary to eliminate the background. The basic schematic of our experiment is outlined in Fig. 1a. We collect PL spatial maps, spectra and transient dynamics of single-domain $t$BLG over a 0.3-1.8 eV laser tuning range. Both acquire PL maps and PLE spectra, we slowly raster scan the diffraction-limited laser over $t$BLG domains in a confocal scanning microscope while collecting the filtered 2–photon PL with back-illuminated EMCCD camera.

Figures 1c plots $\theta$-tunable PL collected upon 2-photon excitation of $t$BLG interlayer electronic states after substraction of the non-resonant background signals such as hot electron emission. The gaussian fits of the PLE spectrum (*solid lines*) pinpoint the PLE peak energies at $X_A \cong 2.61$, 1.73, and 1.67 eV for the 16.3°, 11.5° and 11.2° domains respectively. Like the PLE spectra (*circles*), the corresponding linear absorption spectra (*dashed lines*) also tune with monotonically with stacking angle θ. The absorption spectra, $\sigma_{tBLG} - 2\sigma_G$ and stacking angles of individual *t*BLG domains were assigned by 1-photon linear absorption microscopy after subtraction of the non-resonant graphene background ($2\sigma_G$).[5]. The energy splitting δ, between the 2-photon PLE peak and the 1-photon linear absorption peaks give the energies that range from δ=160 to 380 meV with θ (see table 1.0).

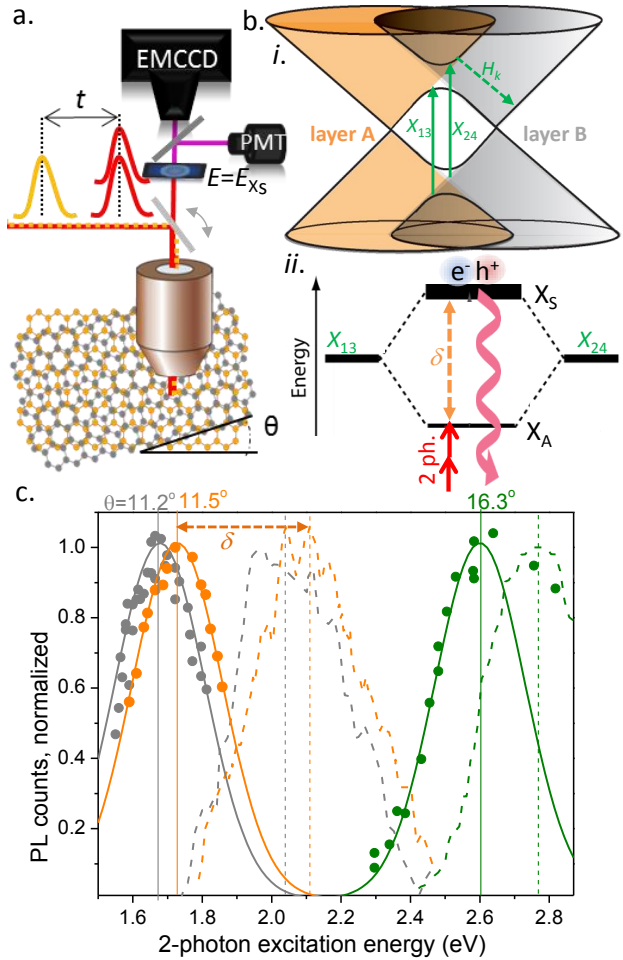

FIG. 1. **θ-tunable photoluminescence. (a)** *Methods*: 2-photon PL, and transient intraband absorption microscopy of single-domain *t*BLG. **(b)** (*i.*) Interband anticrossing regions give degenerate interlayer vHS transitions, $X_{13}$ and $X_{24}$. (*ii.*) A rehybridized model instead predicts symmetric, bright ($X_S$) and antisymmetric, dark ($X_A$) bound-exciton states. **(c)** 2-photon PLE spectra (*circles*, gaussian fits) at θ=11.2°, 11.5° and 16.3° suggest a dark state δ-below the *t*BLG 1-photon linear absorption spectra, (*dashed lines*, $\sigma_{tBLG} - 2\sigma_G$).

The PL and absorption data shown in Fig. 1c, were all collected from spatial maps tuned hyperspectrally for the 1- and 2-photon response. In Fig. 2a, the optical contrast in our hyperspectral linear absorption map show only the 18.3° domain are resonantly excited[5]. Figure 2b shows a representative *t*BLG spatial PL map, collected using a 1.26 eV, 2-photon excitation energy of ∼ $\frac{1}{2}E_{X_A}$. All light emitted below 2.60 eV was removed by a filter stack of OD >6. Strikingly, the two 17.5° domains show localized enhanced PL that is 4-5× stronger than the non-resonant background provided by the surrounding domains that are excited off-resonant. Selective band-pass optical filtering shows the emission energy matches the 1-photon absorption $X_S$ resonance of the 17.5° domain, suggesting the weak resonant emission is thermalized with the dark $X_A$ state, as sketched in Fig. 3a.

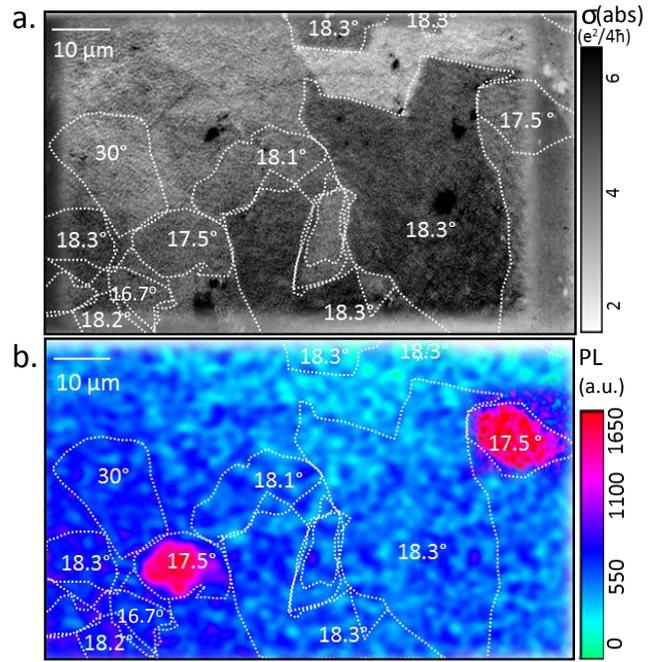

FIG. 2. **Spatial absorption and PL maps. (a)** Linear absorption map showing various angle-stacking domains of an artificially stacked bilayer graphene. Here at 2.9 eV, only the 18.3° domains are resonantly excited. **(b)** Spatial PL map of *t*BLG for resonant 2-photon excitation ($2 \times 1.26$ eV) of the 17.5° $X_A$ state. Enhanced PL is observed only for domains near θ = 17.5°. Other domains show only hot-electron emission, as no exciton states were resonantly excited.

To verify that the detected signal is θ-tunable PL, the possibility of resonantly-enhanced 2-photon scattering processes like SHG or 2-photon resonant Raman must be excluded. While 1-photon resonant enhanced Raman is documented from *t*BLG for a G-band splitting energy of ∼0.2 eV,[5] this interpretation requires a θ-independent red-shift (δ, Stokes) and blue-shift (Δ, antiStokes).[16–18] However, our peak splitting energies vary strongly with θ from δ=0.16 to 0.38 eV, ruling out any 2-photon of resonant Raman enhancement mechanism. Hot electron PL is ruled as this post-thermalization process give only non-resonant, broad emission, whereas we report a 4-5x PL enhancement over the background bilayer signal for the resonantly excited domains shown in Fig. 2b and 3b.

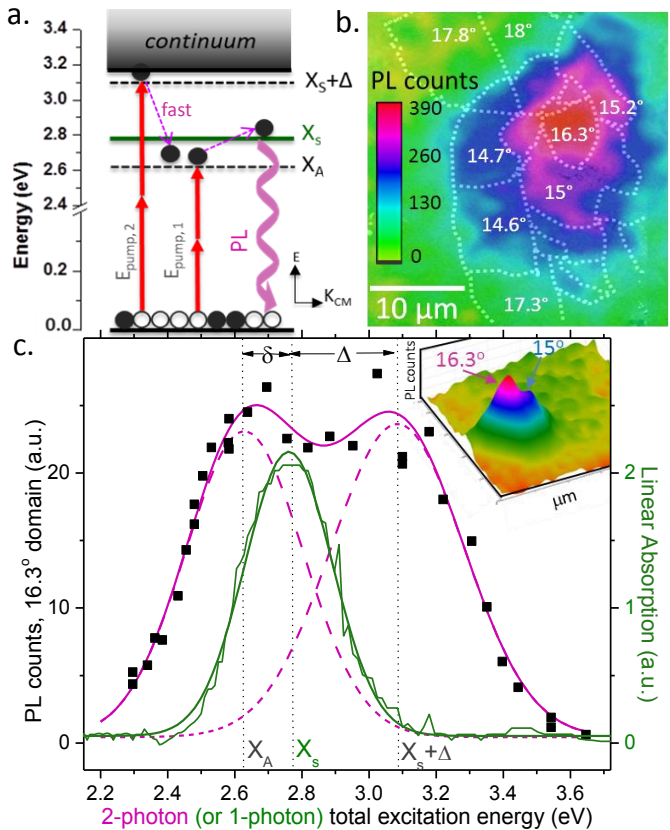

**FIG. 3. 1- and 2-photon interlayer optical transitions.** (a) Bound-exciton model diagram for $\theta = 16.3^o$ sketches emission from the $X_S$ state after two-photon excitation of the $X_A$ state. (b) Spatial PL map of CVD $t$BLG show enhanced emission from the $16.3^o$ and $15.2^o$ domains upon 2-photon excitation of $X_A$ at $2 \times 1.3$ eV. (c) 2-photon PLE spectrum ($black$) fits to a bimodal gaussian about the 1-photon linear absorption ($green$). The total peak splitting $\delta + \Delta$, suggests $E_b \cong 0.5$ eV. (inset) 3D PL maps of grain.

Both the resonant $t$BLG interlayer 2-photon PL and TA signal intensity versus pump fluence exhibits a linear behavior in stark contrast to 1-photon square root dependence measured concurrently. This dependence is consistent with Auger mutiexciton interactions previously documented (i.e. $I_{PL}(n) \propto n_{2ph}^2 \sqrt{n_{aug}} \propto n$), and seen in other strongly-bound exciton system such as semiconducting carbon nanotubes (CNTs).[19–24] As a control, we recovered the conventional quadratic pump power dependence for 2-photon PL excitation of CdSe quantum dots under the same experiment conditions and verified the 1-photon TA signal $t$BLG had a square-root dependence (*see supplemental information*). Previously reported TA on both CNTs and $t$BLG have pump-power responses that show analogous linear and square-root dependencies for 2-photon and 1-photon excitations.[19, 25]

Employing both 2-photon PL and TA intraband methods, we obtain multiple first estimates of the $t$BLG exciton binding energy, $E_b$.[11, 12] Figure 3a extends our simplified $t$BLG exciton model sketched in Fig. 1b$ii$, by adding a manifold of 2-photon states at an energy, $\Delta$ above the absorption resonance. Upon 2-photon excitation of the higher lying dark state at $E_{pump2}$, we find the carriers relax impulsively to the lower lying dark state $X_A$. Since the dark state $X_A$ is a longer-lived state[25], 2-photon excitation at either $E_{pump1}$ or $E_{pump2}$ enables the bright $X_S$ state to emit PL owing to the strong overlap of the $X_A$ and $X_S$ peaks shown in Fig. 3c. This large bright-dark state splitting helps explains why the PL emmisson from $t$BLG is so weak; a similar PL quenching mechanism is thought to exist for semiconducting CNTs.[26]

In Figure 3b, a spatial PL map of $t$BLG domains grown directly by CVD and transferred to silicon nitride show enhanced PL localized to the $16.3^o$ domain whereas the surrounding domains are off-resonant. The similarly stacked $15.2^o$ domain also emits because of overlapping resonances. In Figure 3c, the PLE spectral fits (*magenta*) for the $16.3^o$ domain shows a bimodal distribution with two peaks spanning the 2.76 eV (*green*) absorption resonance at 2.61 eV and 3.1 eV. We obtain a new splitting energy, $\Delta \sim 0.34$ eV between the absorption resonance, $X_S$ and the higher energy PL emission, $X_S + \Delta$. This 2-photon peak is almost twice as spectrally broad as the $X_S$ or $X_A$ peaks. From this, we estimate a lower-bound for the exciton binding energy from the 2-photon PLE peak splitting as $E_b \cong \delta + \Delta = 0.5$ eV for a $16.3^o$ domain. We summarize the $\theta$−dependent peak splitting in the table 1.0 below:

| $\theta$ | $\delta$(eV) | $X_S$(eV) | $\Delta$(eV) | $E_b$ (eV) | figure ref. |
|---|---|---|---|---|---|
| $16.3^0$ | 0.16 | 2.75 | 0.34 | 0.50 | 3b (PL) |
| $11.5^0$ | 0.37 | 2.10 | - | >0.5 | 1c (PL) |
| $11.2^0$ | 0.38 | 2.05 | - | >0.5 | 1c (PL) |
| $7.9^0$ | 0.34 | 1.55 | 0.33 | 0.69 | 4b (TA, blue) |
| $7.9^0$ | – | 1.55 | - | 0.70 | 4b (TA, gray) |

**Table 1.0** - Estimates of the exciton peak splitting of the lower $X_A = X_S - \delta$ and upper $X_S + \Delta$ states. Binding energy estimate assume $E_b \gtrsim \delta + \Delta$. Three independent methods were used; 2-photon PLE, 2-photon TA spectra and 1-photon intraband ESA spectra.

**DISCUSSION:** In optical scattering experiments on $t$BLG, no excitons are excited. As such, the hot electron vHS model in Fig. 1b$i$ sufficiently predicts previously reported $t$BLG resonant enhancements such as Raman $G$-band peaks, STM, ARPES and circular dichroism experiments [1–3, 27–29]. For optical absorption, strong electron-hole interactions are well-documented even in single-layer graphene which has a ∼0.4 eV red-shift renormalization of the optical spectrum near the $M$ saddle-point.[30] Likewise, $0^o$ Bernal stacked bilayer graphene has bound excitons under a high-field.[31] In 1D metallic carbon nanotubes, strong quantum confinement also gives bound excitons with $E_b \cong 0.05$ meV. [32, 33] Similar to semiconducting carbon nanotubes, we find $t$BLG also have low-lying dark states that result in very small PL quantum yield.[26, 33]

Resonant PL emission is not commonly observed in metallic materials because of strong screening of the electron hole pair interaction, and fast thermalization with continuum states. The existence of strongly-bound, stable interlayer excitons under ambient conditions is controversial as the tight-binding model $t$BLG band structure (Fig. 1$bi$) only has avoided crossing regions, no band gap. Previous evidence supporting the bound-exciton model comes primarily from *ab initio* calculations and experimental TA studies showing a population relaxation bottleneck after resonant excitation of $t$BLG.[7, 25] Specifically, Liang *et al.* proposed a strongly-bound interlayer exciton state, $X_A$ calculated to have no electronic coupling with the metallic graphene continuum states ($H_k \cong 0$, see Fig. 1b).[7] Historically, such dark exciton states have been termed ghost Fano resonances.[7, 25]

In Fig. 4, the ultrafast spectral dynamics of interlayer carriers in $7.9^o$ $t$BLG are mapped spectrally and temporally by raster scanning a collinear pump and probe beam. We measure the differential TA of the probe pulse and plot the transient spectra at $t=0.5$ ps in Fig. 4b, and the relaxation kinetics in Fig. 4c. To measure intraband transitions (dashed blue arrows in Fig.4a), the pump beam was resonant at $E_{X_A}=1.54$ eV and near-IR probe energies were scanned from 0.4 eV-1.2 eV to observe the promotion of carriers to the higher lying states (e.g. $X_S + \Delta$) as an excited state absorption (ESA). To isolate dynamics intrinsic to the interlayer electronics, the graphene $0^o$ bilayer TA contribution ($\Delta\sigma_B$) is subtracted from the tBLG TA response ($\Delta\sigma_T$), at each probe energy and plotted in Figs. 4b and c.

Using intraband ESA transient absorption microscopy to relax 2-photon interband selection rules, Fig. 4 provides a complimentary method to our 2-photon estimates of $E_b$. It has been shown for SWCNTs and TMDs, that the exciton binding energy can be estimated by an ESA spectrum between intraband states, where the selection rules for optically dark transition are relaxed in a hydrogenic exciton model (see Fig. 4a, blue arrow).[34] In Fig. 4b, the corresponding spectrum (*blue*) peaked at ∼0.7 eV and corresponds to an ESA response. This suggests an intraband transition to quasi-continuum states depicted in 4a and suggests that $E_b$ ∼0.7 eV for $\theta = 7.9^o$ $t$BLG.

We independently verify the above intraband-transition estimate of $E_b$ by also using 2-photon TA microscopy(see Fig. 4, all data *in grey*). Specifically, by scanning the 2-photon pump energy over across the $X_A$ and $X_A + \Delta$ states we can probe the transient interband optical conductivity of the bright state $X_S$. The resulting 2-photon absorption spectrum shown in Figure 4b (*gray*). We show two enhanced ground state bleach responses peaked at 1.18 and 1.82 eV (*gray*) spanning the 1-photon absorption resonance at 1.54 eV (*green*). We estimate the binding energy by taking: $\delta + \Delta \cong 0.69$ meV. Both of $E_b$ measurements; the near-IR intraband TA, and 2-ph TA microscopy of $7.9^o$ $t$BLG domains independently report the same value or $E_b$ of $t$BLG. Furthermore, our $\theta$-tunable exciton binding energies in Table 1.0 of $E_b \cong 0.5$-0.7 eV agree with the theoretically simulated value of 0.7 eV binding energy for $21^o$ $t$BLG domain.[7]

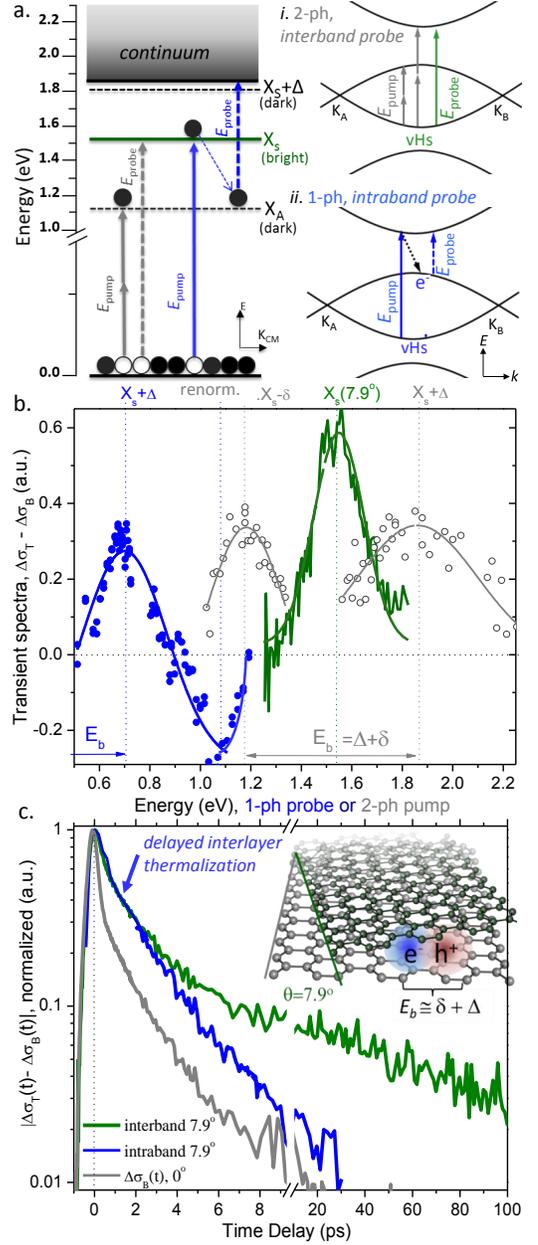

FIG. 4. $E_b$ **from intraband response (a)** Two TA experiments, (*blue*): Fixed pump $X_S$, then probe intraband ESA spectrum near continuum states. (*grey*): 2-photon pump, with fixed probe at $X_S$. (*i.*) and (*ii.*) A TBM vHs depiction of the equivalent TA experiment. **(b)** Corresponding 1- and 2-photon TA spectra, with linear absorption spectrum in green. **(c)** Corresponding interband (*green*) vs. intraband (*blue*)normalized TA kinetics show an absent fast electron thermalization timescale present in $0^o$ bilayer graphene.

Lastly, looking at the charge carrier lifetimes in $t$BLG, we find long-lived intraband response with lifetime of 1.03 ps for the intraband response ($blue$) as shown in Fig. 4c. The resonant interband ($green$) $t$BLG carrier relaxation dynamics have been previously reported.[25] Comparing the two TA decays plotted after subtraction of the much weaker, non-resonant Bernal, $0^o$ graphene response, $\Delta\sigma_B$ ($gray$), we note the absence of the fast electronic thermalization component in $t$BLG. This suggests that the $X_A$ state, relaxes primarily thorough convention graphene phonon assisted processes.[35] Our observations of delayed thermalization of resonantly excited interlayer carriers for both the interband and intraband kinetics supports the theoretical prediction that bound-exciton has weak or vanishing coupling to metallic continuum state ($H_k \cong 0$), stabilizing the interlayer exciton state.[7]

**CONCLUSIONS**: In summary, we report resonant $\theta$-tunable PL from 2-photon resonantly excited $t$BLG and estimate the exciton binding energy by promoting transient carriers to the continuum edge using 1-photon intraband TA. To unambiguously remove contributions from the non-resonant hot electron PL and SHG scattering, we make use of 2-photon excitation and exciton selection rules to observe the week interlayer PL. We observed resonant interlayer PL from both as-grown CVD samples (Figs. 1c and 3) and artificially dry-transferred bilayers (Fig. 2) across many stacking angles ($\theta = 11 - 16.9^o$). We find the PL and lifetime signatures of strongly-bound interlayer excitons are only observed if the exciton state is resonantly prepared by 1- or 2-photon excitation. In particular, we observe resonant PL emission centered at $X_S$ and delayed thermalization rates only for 1-photon excitation of $X_S$, or 2-photon excitations resonant with $X_A$ or $X_S + \Delta$.

Collection of resonant $\theta$-tunable PL is not supported by the vHs $t$BLG electronic model as unbound carriers would thermalize impulsively fast with the metallic continuum states. However, resonantly prepared, stable, strongly-bound interlayer excitons can emit PL as the results of much slower (0.9-1.2 ps) thermalization of the dark strongly bound exciton state, $X_A$ owing to suppressed coupling to the metallic continuum states from the predicted ghost Fano resonance effect. We further estimate the $t$BLG exciton binding energy form the spectral peak splitting we observe in both near-IR intraband transient absorption microscopy and 2-photon PLE microscopy. As summarized in Table 1.0, our estimates of the exciton binding energy changes appreciably with $\theta$ increasing from $\sim 0.4$ eV at $16.3^o$ to $\sim 0.7$ eV at $7.8^o$. Such large values from $E_b$ are comparable to other 2D excitonic materials like TMDs and larger than the measured binding energy of 0.3-0.5 eV in semiconductor SWCNTs, 50 meV in metallic CNTS. [33, 36–39] Our results support that $t$BLG is a first novel 2D hybrid material with stabilized bound excitons that can coexist alongside metallic continuum states for timescales of $\sim 1$ ps. The dual metal-exciton nature of optically excited twisted bilayer graphene suggest new materials applications for efficient light harvesting technology and fast optoelectronics devices.

**Methods summary:** $t$BLG was prepared both directly by chemical vapor deposition (CVD) and manually-stacking layer by the dry transfer method.[40, 41] Substrates were silicon nitride and fused silica and measurements were generally performed in an ambient under continual nitrogen purge. The 2-photon PL results were invariant to the stacking method and substrates used. The stacking angles of individual $t$BLG domains were assigned primarily with hyperspectral linear absorption microscopy after subtraction of the non-resonance graphene background ($\sigma_G$).[5]. $\theta$-assignments were later confirmed with either transient absorption (TA) or dark field TEM [4, 6, 25].

For the interband ESA and 2-photon photoluminescence spectral and ultrafast measurements, a Coherent Chameleon oscillator (80 MHz, 130 fs) pumping an APE Compact OPO with wavelength range 680 - 4000 nm and a NKT Whitelase supercontinuum fiber. ESA TA experiments to probe intraband transitions (Fig. 4, $blue$) were done using confocal transient absorption microscopy. The 2-photon pump beam was modulated at 1 MHz with an AO-modulator. For the pump power dependence measurements, the probe power was fixed at $\sim 2$ x$10^{12}$ photons/cm². 2-photon pump fluences were on average at $\sim 1$ x $10^{14}$ photons/cm² with fluence-dependence conducted to ensure the sample remained undamaged throughout. The spot size of the pump and the probe beams were measured to be 1.5 $\mu$m after a 50x-IR Olympus objective, or a 52x reflective Cassegrain objective for IR excitation. Rigorous power normalization curves were taken $in$ $situ$ for all spectral dependent measurements. This includes the microscope objective transmission corrections, the spectral response of the detection system, and spectral characteristics of the optical filters were taken into account for each wavelength.

For the 2-photon PL microscopy measurements, the pump beam was generated by the OPO. The beam is then raster scanned by the piezo-scanning mirror and the 1-photon back reflection off the sample on a InGaAs detector provided a way to to map the 2-photon and 1-photon responses concurrently. Long pass optical filters were used in the line before the microscope to block possible SHG light from the laser source. The emitted photoluminescence was measured with a TE cooled, back-illuminated EMCCD camera (ProEm HS,

Princeton Instruments, 95% QE) and Hamamatsu Si PMT was also used a secondary detection confirmation (with lock-in). To exclude scattered laser light from Rayleigh, Raman and residual second harmonic generation (SHG), collection of photons with energy less than twice the 2-photon excitation energy were excluded, Specific Chroma and Thorlabs short pass and band pass optical filters stacks were used in front of the camera for emission detection to maintain a OD>6 blockage throughout. All the measurements were performed at 295 K unless specified in nitrogen purged environment.

---